\documentclass[final,3p,authoryear,twocolumn,times,sort&compress]{elsarticle}

\usepackage{multirow,setspace,times,amssymb,amsmath,graphicx,color,bm,rotating,subfigure,url,epstopdf}
\usepackage[font=small]{caption}
\usepackage{lineno}
\usepackage{natbib}
\usepackage{array}
\usepackage[colorlinks,linkcolor=red,anchorcolor=blue,citecolor=green]{hyperref}
\usepackage{bigstrut}

\usepackage{dcolumn}
\newcolumntype{d}[1]{D{.}{.}{#1}}

\journal{XXX}

\begin{document}

\begin{frontmatter}

\title{A global economic policy uncertainty index from principal component analysis}

\author[CME]{Peng-Fei Dai}
\author[CME,CCSCA]{Xiong Xiong}
\author[BS,SS]{Wei-Xing Zhou \corref{cor1}}
\cortext[cor1]{Corresponding author. Address: 130 Meilong Road, P.O. Box 114, School of Business, East China University of Science and Technology, Shanghai 200237, China, Phone: +86 21 64250053, Fax: +86 21 64253152.}
\ead{wxzhou@ecust.edu.cn} 
\address[CME]{College of Management and Economics, Tianjin University, Tianjin 300072, China}
\address[CCSCA]{China Center for Social Computing and Analytics, Tianjin University, Tianjin 300072, China}
\address[BS]{Department of Finance, East China University of Science and Technology, Shanghai 200237, China}
\address[SS]{Department of Mathematics, East China University of Science and Technology, Shanghai 200237, China}

\begin{abstract}
This paper constructs a global economic policy uncertainty index through the principal component analysis of the economic policy uncertainty indices for twenty primary economies around the world. We find that the PCA-based global economic policy uncertainty index is a good proxy for the economic policy uncertainty on a global scale, which is quite consistent with the GDP-weighted global economic policy uncertainty index. The PCA-based economic policy uncertainty index is found to be positively related with the volatility and correlation of the global financial market, which indicates that the stocks are more volatile and correlated when the global economic policy uncertainty is higher. The PCA-based global economic policy uncertainty index performs slightly better because the relationship between the PCA-based uncertainty and market volatility and correlation is more significant.
\end{abstract}


\end{frontmatter}


\section{Introduction}

The study on uncertainty has attracted much attention \citep{Bloom-2009-Em}. \cite{Veronesi-Pastor-2012-JF} and \cite{Veronesi-Pastor-2013-JFE} develop a general equilibrium model to study how changes in government policy choice affect stock prices and explore the relationship between political uncertainty and stock risk premium. \cite{Baker-Bloom-Davis-2016-QJE} construct an index as the proxy for economic policy uncertainty (EPU) in the United States and 11 other major economies, which was initially put forward by \cite{Baker-Bloom-Davis-2013-CBRP}. Many scholars, such as \cite{Moore-2017-ER} and \cite{Arbatli-Davis-Ito-Miake-Saito-2017-IMF}, construct other indices for different economies successively using the same method. \cite{Bontempi-Golinelli-Squadrani-2016} introduce a new uncertainty indicator based on Internet searches. \cite{Castelnuovo-Tran-2017-EL} develop uncertainty indices for the United States and Australia, which are based on Google Trends data. 

Many papers have studied the influence of economic policy uncertainty on the international financial markets. \cite{Li-Zhang-Gao-2015-EL} investigate the impacts of economic policy uncertainty shocks on stock-bond correlations for the financial market in United States. \cite{Kloner-Sekkel-2014-EL} discuss international spillovers of policy uncertainty using the EPU indices from six developed economies. \cite{Brogaard-Detzel-2015-MS} use a search-based measure to capture economic policy uncertainty for 21 economies, and found economic policy uncertainty has a significant effect on the contemporaneous market returns and volatility. 

Recently, the aggregate global economic policy uncertainty (GEPU) has been proposed and investigated. \cite{Davis-2016-NBER} constructs an index of global economic policy uncertainty which is a GDP-weighted average of national EPU indices for 20 economies. \cite{Fang-Chen-Yu-Qian-2018-JFinM} examine whether the GDP-based GEPU index provides predictability for the gold futures market volatility. \cite{Ersan-Akron-Demir-2019-TE} access the effect of the GDP-based GEPU index on the stock returns of travel and leisure companies.

For a financial market with multiple assets, the largest eigenvalue of the correlation matrix of returns, when normalized by the number of assets, quantifies the systemic risk of the market, while its eigenvector reflects the whole movement of the market \citep{Plerou-Gopikrishnan-Rosenow-Amaral-Guhr-Stanley-2002-PRE,Shapira-Kenett-BenJacob-2009-EPJB,Kenett-Tumminello-Madi-GurGershgoren-Mantegna-BenJacob-2010-PLoS1,Song-Tumminello-Zhou-Mantegna-2011-PRE,Kritzman-Li-Page-Rigobon-2011-JPM,Billio-Getmansky-Lo-Pelizzon-2012-JFE,Meng-Xie-Jiang-Podobnik-Zhou-Stanley-2014-SR,Dai-Xie-Jiang-Jiang-Zhou-2016-EmpE,Han-Xie-Xiong-Zhang-Zhou-2017-FNL,Sandoval-2017-JNTF,EmmertStreib-Musa-Baltakys-Kanniainen-Tripathi-YliHarja-Joblbauer-Dehmer-2018-JNTF}.
Inspired by these studies, we construct an alternative index for the aggregate global economic policy uncertainty based on the principal component analysis. In addition, we explore the effect of global economic policy uncertainty on the volatility and correlation of the global stock market.

The remainder of this paper is organized as follows. Section~\ref{S2:Data} presents a brief description of the data. Section~\ref{S3:Method} focuses on the methodology. Section~\ref{S4:Results} documents our findings. Section~\ref{S5:Conclusions} concludes.

\section{Data}
\label{S2:Data}

\begin{figure}[htbp]
\centering
\includegraphics[width=0.96\linewidth]{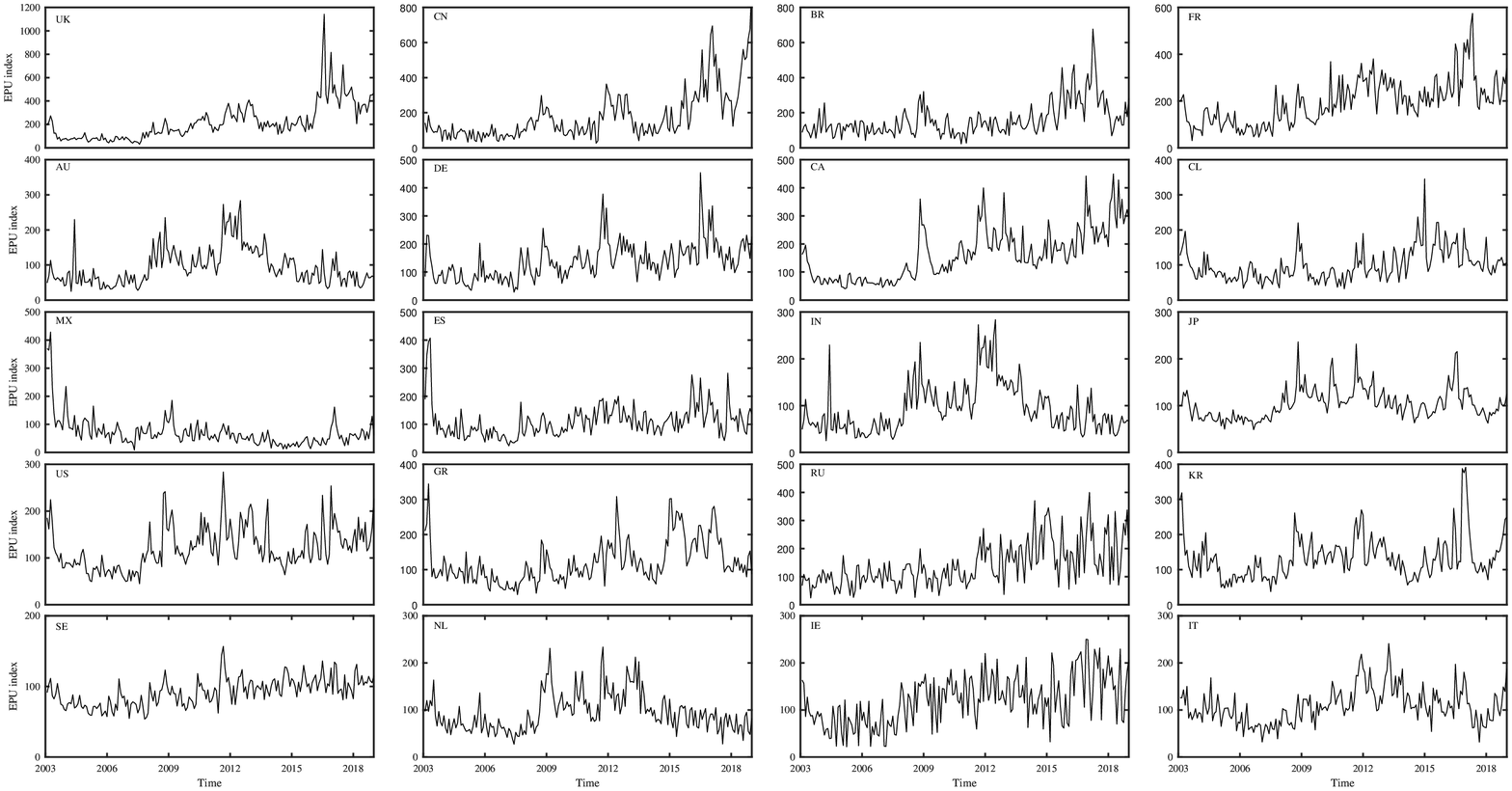}
\caption{The economic policy uncertainty indices  from January 2003 to December 2018 of twenty economies: Australia (AU), Brazil (BR), Canada (CA), Chile (CL), China (CN), France (FR), Germany (DE), Greece (GR), India (IN), Ireland (IE), Italy (IT), Japan (JP), Mexico (MX), the Netherlands (NL), Russia (RU), South Korea (KR), Spain (ES), Sweden (SE), the United Kingdom (UK), and the United States (US).}
\label{Fig:GEPU:Data}
\end{figure}

In order to calculate the PCA-based GEPU index, we retrieve the economic policy uncertainty indices for twenty economies from \url{http://www.policyuncertainty.com}. These twenty economies are Australia (AU), Brazil (BR), Canada (CA), Chile (CL), China (CN), France (FR), Germany (DE), Greece (GR), India (IN), Ireland (IE), Italy (IT), Japan (JP), Mexico (MX), the Netherlands (NL), Russia (RU), South Korea (KR), Spain (ES), Sweden (SE), the United Kingdom (UK), and the United States (US), which are in perfect accordance with the economies that \cite{Davis-2016-NBER} apply to construct the GDP-based GEPU index. Fig.~\ref{Fig:GEPU:Data} shows the evolution of these EPU indices from January 2003 to December 2018, where each index includes 192 monthly observations.

\begin{figure}[htbp]
\centering
\includegraphics[width=0.8\linewidth]{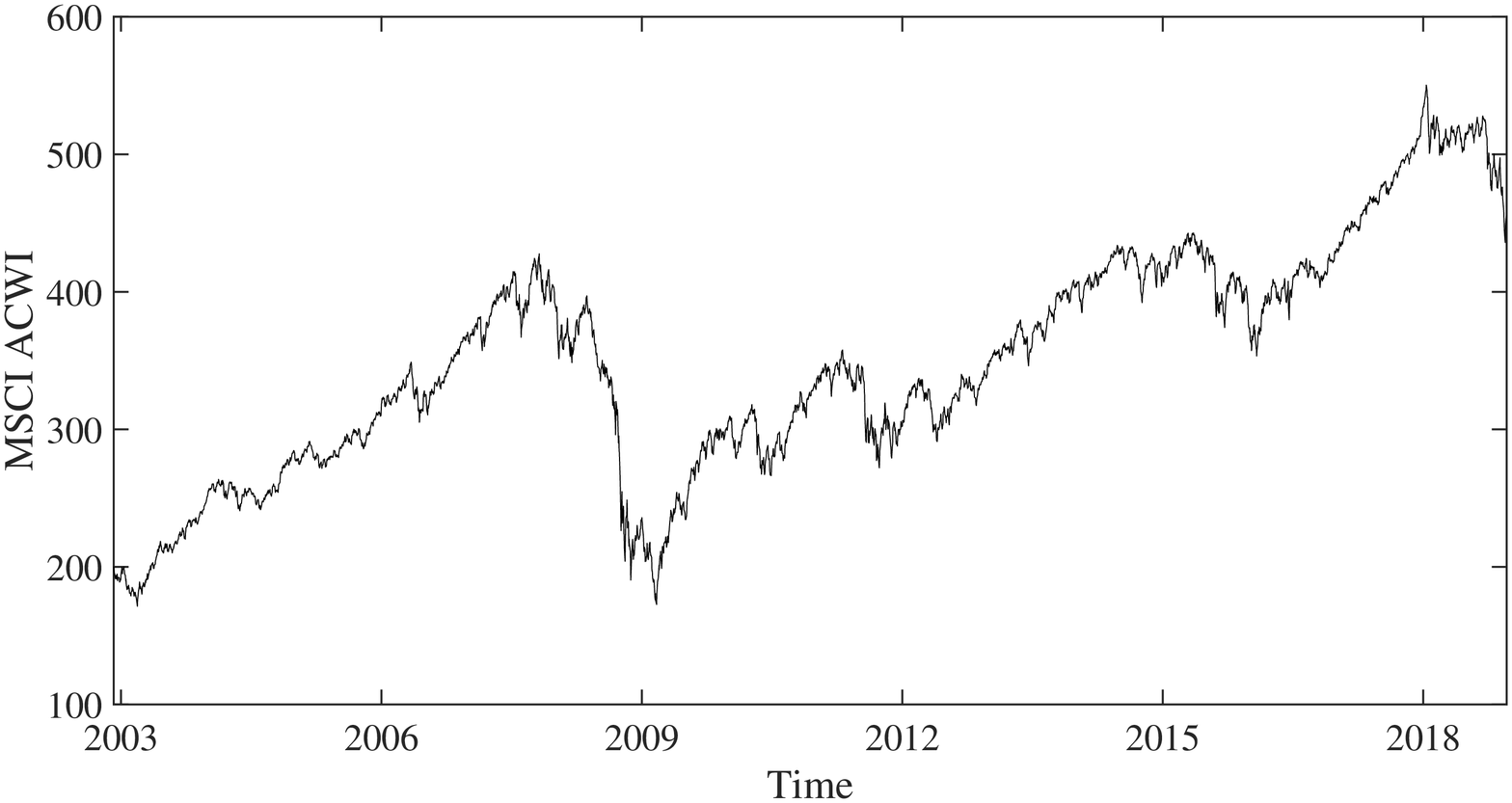}
\caption{The price trajectory of MSCI's All Country World Index from January 2003 to December 2018.}
\label{Fig:MSCI}
\end{figure}

In terms of the global financial market, we select MSCI's All Country World Index (ACWI) as its measure. The MSCI ACWI represents the performance of the stocks across 23 developed and 24 emerging markets. 
Fig.~\ref{Fig:MSCI} displays the trend of MSCI ACWI, which covers the daily closing prices from December 2002 to December 2018. The composite indices for each market (from Bloomberg) are utilized to evaluate the correlations between markets also using the daily closing prices from December 2002 to December 2018.

\section{Methodology}
\label{S3:Method}


First, we normalize the EPU index $i$ for each economy over a moving window $[t-T+1,t]$ of size $T$:
\begin{equation}
\label{Eq:EPUNorm}
x_{i}(s) \equiv ({EPU_i(s)-\left\langle EPU_i(s) \right\rangle_s})/{\sigma_i},
\end{equation}
where $EPU_i(s)$ represents the index for the $i$th economy in the $s$-th month and $\sigma_i$
is the standard deviation of $EPU_{is}$. Second, we obtain the cross-correlation matrix $C$ by computing the pairwise cross-correlation coefficient between any two EPU indices for the economies:
\begin{equation}
\label{Eq:Cij}
C_{ij} \equiv  \left\langle x_i (s)x_j (s) \right\rangle.
\end{equation}
By definition, the elements $C_{ij}$ vary from $-1$ to $1$, where $C_{ij}=1$ corresponds to a perfect positive cross-correlation, $C_{ij}=1$ corresponds to a perfect anti-correlation, and $C_{ij}=0$ reflects no cross-correlations between the indices for economy $i$ and economy $j$. The cross-correlation matrix can also be expressed in the matrix form:
\begin{equation}
\label{Eq:CMatrix}
\bm{{\rm C}}=\frac{1}{T}\bm{{\rm X}}\bm{{\rm{X}}}',
\end{equation}
where $\bm{{\rm{X}}}$ is an $N \times T$ matrix with elements $\{x_i(s): i=1, \ldots, N; s=1, \ldots, T\}$ and $\bm{{\rm{X}}}'$ denotes the transpose of $\bm{{\rm{X}}}$. Third, we obtain the eigenvector $\bm{{\rm u}}_1(t)$ for the largest eigenvalue $\lambda_1(t)$ of the cross-correlation matrix in the $t$-th window:
\begin{equation}
\label{lambda1}
\bm{{\rm C}} \bm{{\rm u}}_1(t)=\lambda_1(t) \bm{{\rm u}}_1(t),
\end{equation}
where $\bm{{\rm u}}_1(t)=\left[u_{11}(t),u_{12}(t),\ldots,u_{1N}(t) \right]$. Finally, we construct the PCA-based GEPU index by the eigenportfolio of the economic policy uncertainty indices for the $N$ economies:
\begin{equation}
\label{Eq:GEPU}
GEPU(t)=\frac{\bm{{\rm u}}_1(t)\cdot\bm{{\rm EPU}}(t)}{\sum_{i=1}^N u_{1i}(t)},
\end{equation}
%
where 
$\bm{{\rm EPU}}(t)=\left[ EPU_1(t), EPU_2(t), \ldots,EPU_N(t)\right]'$.

\section{Empirical results}
\label{S4:Results}

\begin{figure}[htbp]
\centering
\includegraphics[width=0.96\linewidth]{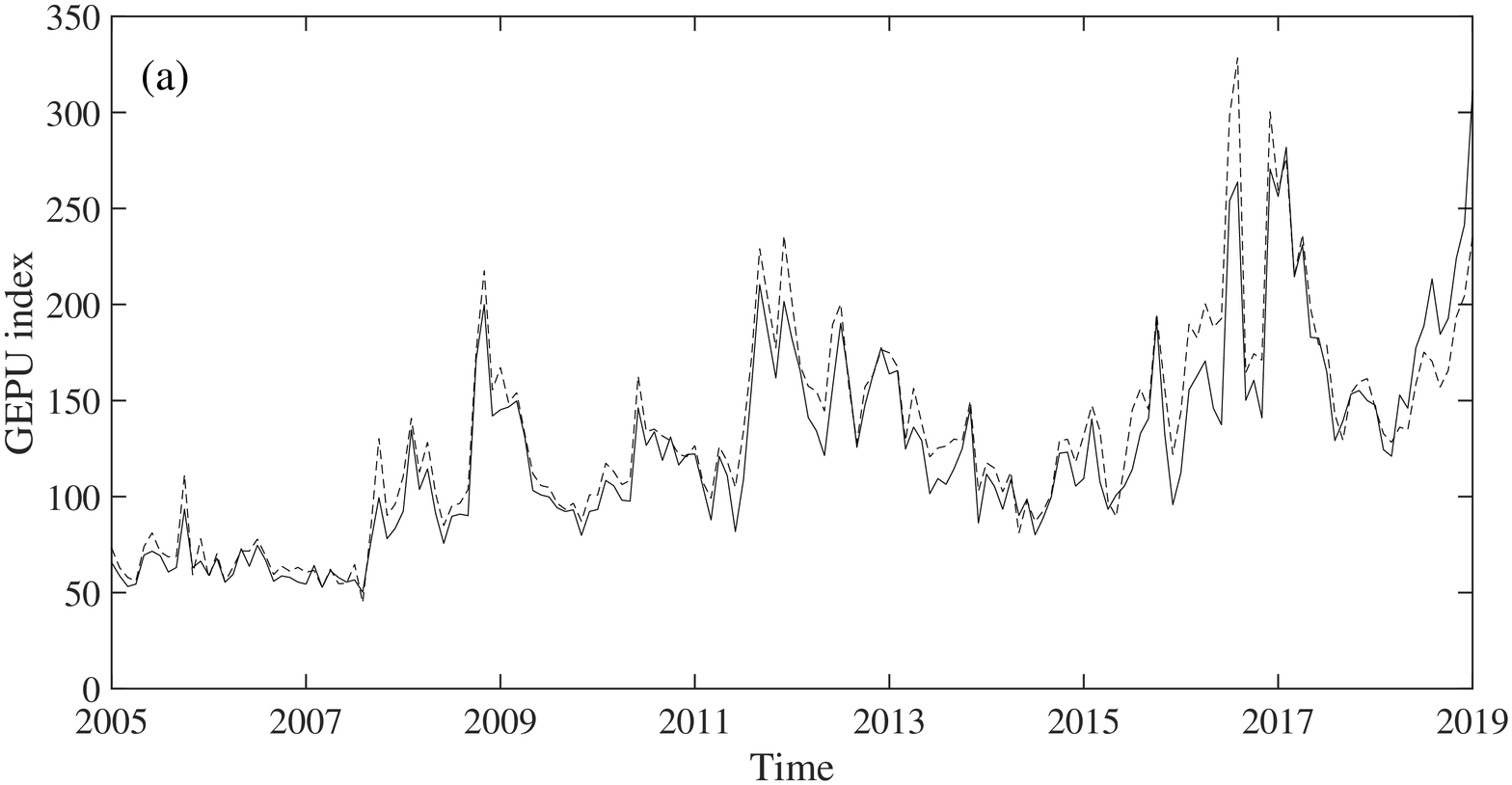}\\
\includegraphics[width=0.96\linewidth]{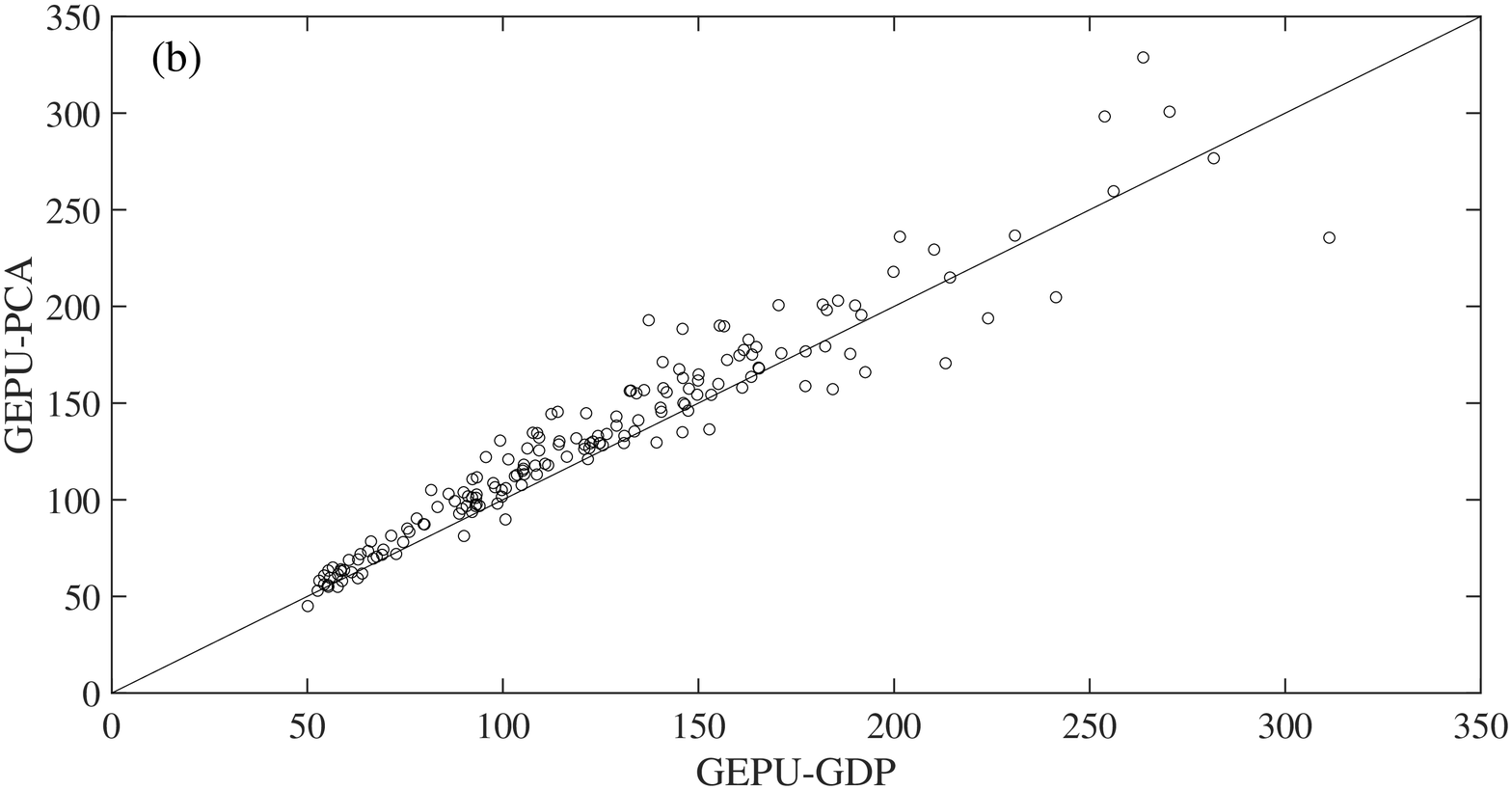}
\caption{Comparison between GEPU-PCA and GEPU-GDP. (a) Evolutionary trajectories of GEPU-GDP (solid line) and GEPU-PCA (dashed line). (b) Scatter plot of the two GEPU indices.}
\label{Fig:GEPU:Comparison}
\end{figure}

In order to examine the robustness of the results, we construct the PCA-based GEPU index for five window sizes $T=24, 30, 36, 42$, and $48$ months.
Fig.~\ref{Fig:GEPU:Comparison} displays the comparison between the GEPU-PCA and GEPU-GDP indices for $T=24$ months. We find that the evolutionary trajectories of the GEPU-GDP and GEPU-PCA indices are close to each other, which is also indicated by the nice linearity of the data points in the corresponding scatter plot. The results for other window sizes are very similar. Overall, all the GEPU-PCA indices are very close to the GEPU-GDP indices, although GEPU-PCA is obtained without using any other economic data. The discrepancy between the two indices increases when the uncertainty is high.

Table~\ref{TB:GEPU:Comparison} shows the correlation coefficients between GEPU-PCA and GEPU-GDP for different window sizes $T$, all of which are greater than 0.94. The window size seems to have no impact on the correlation.

\begin{table}[htp]
\centering
\setlength{\abovecaptionskip}{0pt}
\setlength{\belowcaptionskip}{10pt}
\caption{Correlation between GEPU-PCA and GEPU-GDP for different window size $T$.}
\label{TB:GEPU:Comparison}
\smallskip
\begin{tabular}{cccc}
\hline  \noalign{\smallskip}
    $T$        &   $t_0$     &   Obs.       &  Correlation   \\ 
\noalign{\smallskip}\hline
   24 M      &   2004.12           &   169        &  0.9572       \\
   30 M      &   2005.06           &   163        &  0.9521       \\
   36 M      &   2005.12           &   157        &  0.9417       \\
   42 M      &   2006.06           &   151        &  0.9473       \\
   48 M      &   2007.12           &   145        &  0.9525      \\ \hline
\end{tabular}
\end{table}

\begin{figure}[h!]
\centering
\includegraphics[width=0.96\linewidth]{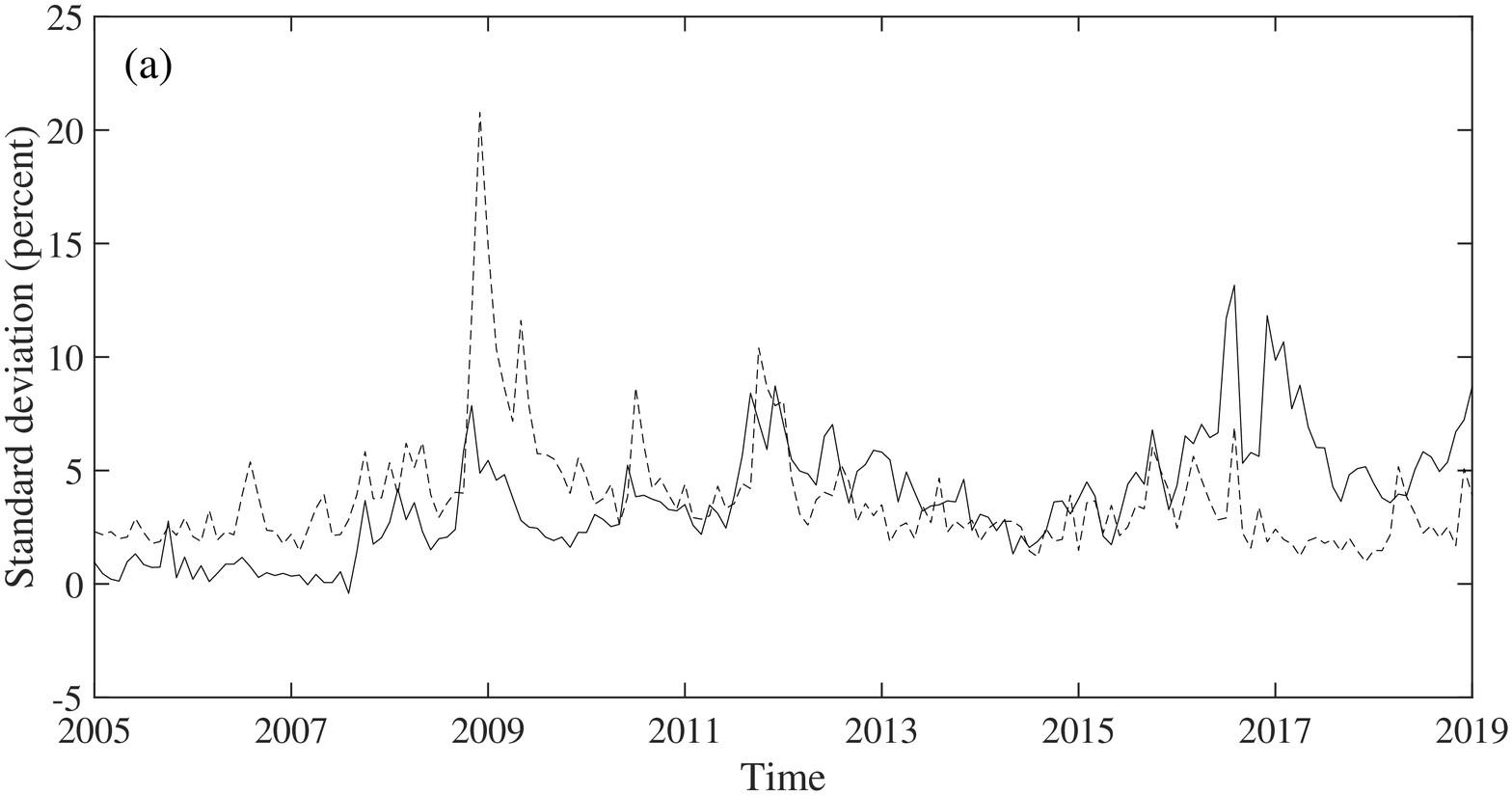}\\
\includegraphics[width=0.96\linewidth]{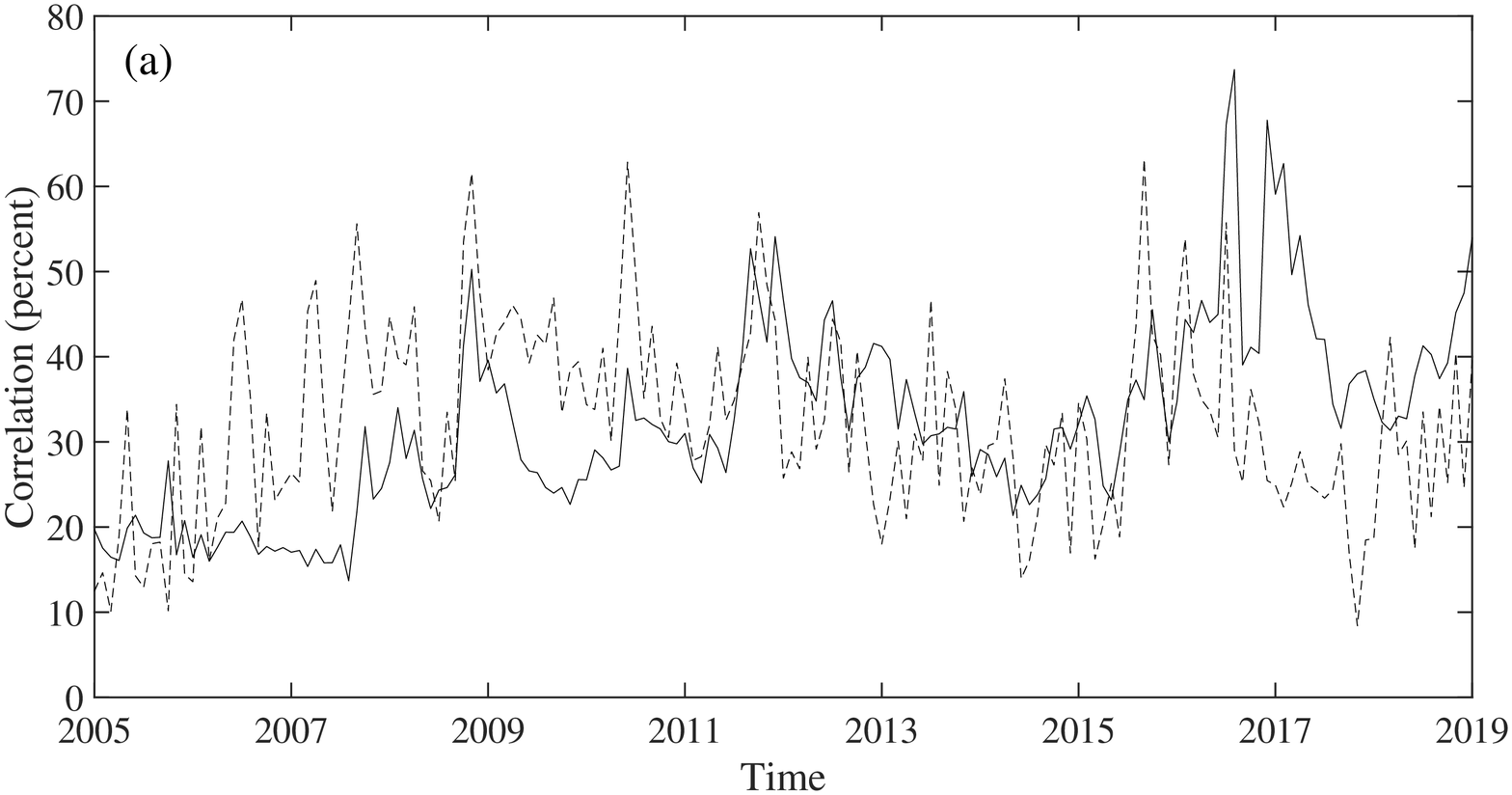}
\caption{GEPU-PCA versus global stock market volitility and correlation. The solid line in each panel plots GEPU-PCA, which is scaled to have the same mean and variance as the other variable plotted in the same panel. The other variable is volatility of MSCI ACWI in the upper panel and the equal-weighted average of pairwise correlation for all stock comprehensive indices corresponding to the 47 economies which are from MSCI ACWI in the lower panel. Both volatility and correlation are calculated monthly from daily returns within the month.}
\label{Fig:GEPU:Vol&Correlations}	
\end{figure}

\cite{Veronesi-Pastor-2013-JFE} provide the theoretical foundation for the positive relations between economic policy uncertainty and both volatility and correlation and conduct an empirical analysis in the American market with the results supporting the theoretical arguments. To verify that this association does also exist in the global market, GEPU-PCA and GEPU-GDP are respectively used as proxy for global economic policy uncertainty. Fig.~\ref{Fig:GEPU:Vol&Correlations} reveals the strong correlation between GEPU-PCA and the two variables about the global market, especially in the first half of the samples. 

For the sake of a comparative analysis, we adopt the same empirical models as \cite{Veronesi-Pastor-2013-JFE}. For volatility, we employ the following regressions:
\begin{equation}
Vol(t)=\beta_{0}+\beta_{1}GEPU(t)+\varepsilon(t)
\label{Eq:GEPU:Vol:1}
\end{equation}
\begin{equation}
Vol(t)=\beta_{0}+\beta_{1}GEPU(t)+\beta_{2}Vol\left(t-1\right)+\varepsilon(t),
\label{Eq:GEPU:Vol:2}
\end{equation}
where $Vol$ represents the volatility. For correlation, the following regressions are considered:
\begin{equation}
Corr(t)=\beta_{0}+\beta_{1}GEPU(t)+\varepsilon(t)
\label{Eq:GEPU:Corr:1}
\end{equation}
\begin{equation}
Corr(t)=\beta_{0}+\beta_{1}GEPU(t)+\beta_{2}Corr\left(t-1\right)+\varepsilon(t),
\label{Eq:GEPU:Corr:2}
\end{equation}
where $Corr$ stands for the correlation. The lagged terms, $Vol\left(t-1\right)$ and $Corr\left(t-1\right)$, eliminate most of the autocorrelation in the dependent variable series. 

\begin{table}[h!tp]
\centering
\setlength{\abovecaptionskip}{0pt}
\setlength{\belowcaptionskip}{10pt}
\caption{Global economic uncertainty, volatility, and correlation.
}
\label{TB:GEPU:Vol:Corr}
\begin{tabular}{cccccccccc}
\hline
\noalign{\smallskip}
\multicolumn{3}{l} {Panel A: Volatility} \\ \hline
Obs.          &  Eq.~(\ref{Eq:GEPU:Vol:1})  &   Eq.~(\ref{Eq:GEPU:Vol:2})&&  Eq.~(\ref{Eq:GEPU:Vol:1})  &   Eq.~(\ref{Eq:GEPU:Vol:2})        \\\hline
\noalign{\smallskip}
& \multicolumn{2}{c} {GEPU-PCA} && \multicolumn{2}{c} {GEPU-GDP} \\
\cline{2-3} \cline{5-6}
169      &    0.0018        &       0.0010        &&    0.0017        &       0.0011            \\
               &   (3.02)       &      (2.23)       &&   (2.70)       &      (2.17)         \\
163      &    0.0017        &       0.0010        &&    0.0016        &       0.0011      \\
               &   (2.74)       &      (2.09)       &&   (2.50)       &      (2.09)       \\
157      &   0.0015         &       0.0009        &&    0.0016        &       0.0011      \\
               &   (2.36)       &      (1.88)       &&   (2.31)       &      (2.01)      \\
151      &   0.0014         &       0.0009        &&    0.0015        &       0.0011      \\
               &   (2.11)       &      (1.81)       &&   (2.12)       &      (1.95)      \\
145      &   0.0013         &       0.0009        &&    0.0015        &       0.0011      \\
               &   (1.89)       &      (1.71)       &&   (1.96)       &      (1.87)      \\ \hline
\noalign{\smallskip}
\multicolumn{3}{l} {Panel B: Correlation} \\ \hline
Obs.          &  Eq.~(\ref{Eq:GEPU:Corr:1})  &   Eq.~(\ref{Eq:GEPU:Corr:2})&&  Eq.~(\ref{Eq:GEPU:Corr:1})  &   Eq.~(\ref{Eq:GEPU:Corr:2})        \\\hline
\noalign{\smallskip}
& \multicolumn{2}{c} {GEPU-PCA} && \multicolumn{2}{c} {GEPU-GDP} \\
\cline{2-3} \cline{5-6}
169          &   0.0459        &        0.0194      &&    0.0377        &       0.0163       \\
                   &   (2.88)       &       (1.36)     &&   (2.30)       &      (1.14)       \\
163          &   0.0311        &        0.0127      &&    0.0262        &       0.0112       \\
                   &   (1.96)       &       (0.90)     &&   (1.58)       &      (0.77)        \\
157          &   0.0152        &        0.0051      &&    0.0154        &       0.0063         \\
                   &   (0.97)       &       (0.37)     &&   (0.99)       &      (0.43)         \\
151         &   0.0017        &        0.0003      &&    0.0067        &       0.0038         \\
                   &   (0.10)       &       (0.02)     &&   (0.39)       &      (0.25)        \\
145          &  -0.0063        &       -0.0054      &&    0.0020        &      -0.0005         \\
                   &  (-0.04)       &      (-0.04)     &&   (0.11)       &     (-0.03)        \\ \hline
\end{tabular}
\end{table}

Table \ref{TB:GEPU:Vol:Corr} reports the estimates of  $\beta_{1}$ and their $t$-statistics in all the forty regressions. Panel A shows that  $\beta_{1}>0$ in all twenty regressions and all the 20 point estimates are significant at the $10\%$ level, which provides strong evidence for the theoretical foundation that the global market should be more volatile when there is higher economic policy uncertainty. Panel B presents weaker supporting evidence for the associated theoretical foundation since only three point estimates of $\beta_{1}$ are at the $10\%$ level although $\beta_{1}$ is positive in 17 of the 20 regressions. 

We also find that after removing the autocorrelation in the volatility and correlation, the coefficient $\beta_1$ decreases. In addition, in most cases, $\beta_1$ increases with increasing sample length, except for GEPU-GDP versus volatility using Eq.~(\ref{Eq:GEPU:Corr:2}) in which $\beta_1$ is independent of the sample length. Indeed, we argue that this trending phenomenon is caused by the fact that the correlation is stronger between market volatility (or correlation) and uncertainty in early years, as shown in Fig.~\ref{Fig:GEPU:Vol&Correlations}.

\section{Conclusions}
\label{S5:Conclusions}

This paper constructs a novel index of global economic policy uncertainty based on the principal component analysis. This index is shown to be quite close to the GDP-weighted average global economic policy uncertainty index. We employ both GEPU-PCA and GEPU-GDP as the proxies for uncertainty to investigate the association between the global financial market and economic policy uncertainty and find that the global market should be more volatile and correlated when there is higher economic policy uncertainty. Moreover, GEPU-PCA performs lightly better than GEPU-GDP when the observations are enough in the sense that the correlations are more significant when GEPU-PCA is adopted in the analysis.

\section*{Acknowledgements}

This work was supported by the National Natural Science Foundation of China (Grants No. 71532009, U1811462, 71790594), the Fundamental Research Funds for the Central Universities, and Tianjin Development Program for Innovation and Entrepreneurship.


\end{document}